# Straightforward Bias- and Frequency-Dependent Small-Signal Model Extraction for Single-Layer Graphene FETs

Nikolaos Mavredakis, Anibal Pacheco-Sanchez, Wei Wei, Emiliano Pallecchi, Henri Happy, and David Jiménez

*Abstract*—**We propose an explicit small-signal graphene field-effect transistor (GFET) parameter extraction procedure based on a charge-based quasi-static model. The dependence of the small-signal parameters on both gate voltage and frequency is precisely validated by high-frequency (up to 18 GHz) on-wafer measurements from a 300 nm device. These parameters are studied simultaneously, in contrast to other works which focus exclusively on few. Efficient procedures have been applied to GFETs for the first time to remove contact and gate resistances from the *Y*-parameters. The use of these methods yields straightforward equations for extracting the small-signal model parameters, which is extremely useful for radio-frequency circuit design. Furthermore, we show for the first time experimental validation vs. both gate voltage and frequency of the intrinsic GFET non-reciprocal capacitance model. Accurate models are also presented for the gate voltage-dependence of the measured unity-gain and maximum oscillation frequencies as well as of the current and power gains.**

## 1. Introduction

Research on graphene devices (GFETs) is on the rise and prevails the state of the art of RF applications with emerging two-dimensional technologies [1]. Exceptional extrinsic maximum oscillation frequencies ($f_{max}$) above 100 GHz have been reported [2] whereas, for short-channel GFETs with gate lengths in the range of 100-500 nm, $f_{max}$ and extrinsic unity-gain cut-off frequency ($f_{tEXT}$) are commensurate when compared with Si MOSFETs [3 (figure 3d-3e)-4] of similar dimensions [5-6]. Such prominent performance despite the still premature phase of the GFET technology, has driven circuit designers to demonstrate fundamental analog and RF circuits such as mixers [7-8], low noise [9] and power amplifiers [10], frequency multipliers [11], receivers [12] and balun architectures [13]. These RF circuits have been enabled mainly by table-based/empirical models which are of practical use but lack of a correct description of internal device phenomena and hence, the reproducibility and feasibility of such applications might be questionable.

Thus, reliable physics-based transport-RF/small-signal GFET compact models are prerequisite for an adequate design of the aforementioned circuits. An abundant amount of such GFET models has been demonstrated so far [14-24]. Usually, Meyer-like [25] equivalent circuits have been used [14-18, 22-23] which might offer straightforward and fast computations but do not ensure charge conservation in the intrinsic device. A charge-based model, firstly introduced in [19] and afterwards used elsewhere [20-21, 24], takes into consideration the non-reciprocal characteristics of intrinsic capacitances hence, it guarantees charge conservation. All the preceding works however, lack of concurrently validating with measured data, both the bias- and frequency-dependence of most of the small-signal parameters such as intrinsic capacitances $C_{GS}$, $C_{GD}$, $C_{GG}$, $C_{DG}$, $C_{SD}$ (G, S, D are the gate, source and drain of the device, respectively), intrinsic and extrinsic $f_{tINT}$, $f_{tEXT}$ and $f_{max}$, small-signal current and unilateral power gains, $|h_{21}|$ and $U$ as summarized in Table I. A non-quasi-static (NQS) model is proposed in [24] but is not validated with experiments; such NQS effects are out of the scope of the present study. Notice that, even though a charge-based scheme is claimed in [20] and hence, non-reciprocal capacitances are considered, $C_{DG}$ which differ from $C_{GD}$ under such conditions, (in contrast to Meyer-like approaches) is not presented[1].

Hence, the main goal of the current work is to provide an extensive picture of both bias- and frequency-dependent GFET modeling of all the crucial small-signal parameters in comparison with measured data. Our methodology is based on an accurate charge-based model [19, 21] in contrast to most previous works. Our efforts are focused on the quasi-static (QS) regime below $f_{tEXT}$, which is a valid consideration for applications such as RF amplifiers. The implementation of effective procedures for contact [23, 26-27] and gate resistances' [5-6, 28] $R_C$, $R_G$ elimination, permits the extraction of straightforward expressions for all measured

---

N. Mavredakis, A. Pacheco and D. Jiménez are with the Departament d'Enginyeria Electrònica, Escola d'Enginyeria, Universitat Autònoma de Barcelona, Bellaterra 08193, Spain. (e-mail: Nikolaos.mavredakis@uab.es).
W. Wei, E. Pallecchi and H. Happy are with Univ. Lille, CNRS, UMR 8520 - IEMN, F-59000 Lille, France.

[1]Besides, no de-embedding structures are used in [20] and the parasitic elements are extracted through electromagnetic simulations. It is also mentioned in the text that the maximum $f_t$ where they fit their model [20 (figure 5f)] is achieved without de-embedding.

| Validations | Parameters | References |
|---|---|---|
| w/ NDS vs. $V_{GS}$ | All intrinsic capacitances | [19] |
| w/ meas. vs. $V_{GS}$ | $C_{GS}$, $C_{GD}$, $f_t$ | [20] |
| w/ meas. vs. freq | $|h_{21}|$, $U$ | [20] |
| w/ NDS vs. $V_{GS}$ | All intrinsic capacitances, $f_t$, $f_{max}$ | [21] |
| w/ meas. vs. freq at 3 $V_{DS}$ values | $|h_{21}|$, $U$ | [21] |

intrinsic parameters which can be a helpful tool for circuit designers in terms of fast first-order model estimation [18]; this is not the case for more intricate equations derived from complicated $R_C$, $R_G$ removal methods [21 (equations 6-17)].

## 2. DUT and Measurement Setup

The proposed modeling method has been validated with on-wafer DC-RF measurements from a single-layer short-channel aluminum back-gated CVD GFET with a ~4 nm thick Al$_2$O$_3$ used as a dielectric layer between graphene and gate as shown elsewhere [29-30]. The total width is $W$=12x2 μm=24 μm (where 2 is the number of gate fingers) and the length $L$=300 nm. Au source-drain contacts are used which can ensure very low $R_C$.$W$ in range of 125 Ω.μm resulting in higher transconductance $g_m$ and $f_t$, $f_{max}$ [3]. DC measurements have been conducted with an Agilent E5260B parameter analyzer where $V_{DS}$ is set to 0.5 V while $V_{GS}$ is swept from 0 to 0.7 V in the p-type region of the GFET operation. S-parameters have been also measured at the aforementioned bias points with an Agilent E8361A Vector Network Analyzer from 2 GHz up to 18 GHz; an "OPEN" dummy structure, fabricated on the same chip, has been used for de-embedding.

## 3. Parameter Extraction

A schematic cross section of the GFET under test is presented in figure 1a while the charge-based small-signal equivalent circuit used in this study is shown in figure 1b (cf. [31 (figure 8.5)]). Apart from the intrinsic device, source and drain contact resistances $R_S$=$R_D$=$R_C$/2 (regarded equal as in [14-16, 19, 21] which does not affect the calculation of intrinsic parameters since: $V_{DSin}$=$V_{DS}$-$I_D(R_S+R_D)$) and $R_G$ are also considered as well as extrinsic parasitic capacitances $C_{GSP}$, $C_{GDP}$, $C_{SDP}$. The latter are eliminated through an OPEN de-embedding procedure. Thus, device de-embedded $Y$-parameters are given by: $Y_{DEV}$=$Y_{MEAS}$-$Y_{OPEN}$ where $Y_{MEAS}$ are the raw measurements and $Y_{OPEN}$ the OPEN structure measured $Y$-parameters, respectively. To correctly extract the small-signal model, the transport model parameters should be precisely estimated. Those are extracted from the measured $\Re(Y_{21DEV})$ and $\Re(Y_{22DEV})$ which are the extrinsic (after de-embedding but before $R_G$ and $R_C$ removal) transconductance $g_m$ and output conductance $g_{ds}$ of the device, respectively. The extracted parameters are presented in Table II where $\mu$ is the carrier mobility, $C_{back}$ the back-gate

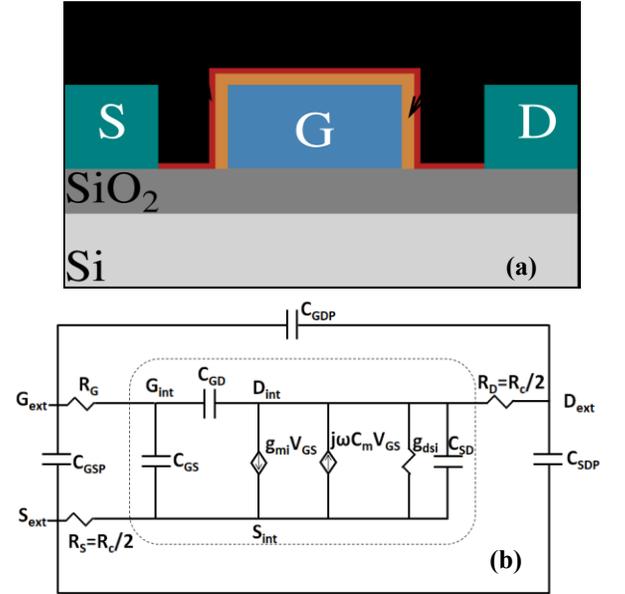

**Figure 1.** (a) Schematic of the GFET under test (b) charge based small-signal equivalent circuit. $g_{mi}$, $g_{dsi}$ are the intrinsic transconductance and output conductance, respectively. $C_m$=$C_{DG}$-$C_{GD}$ where $C_{GS}$, $C_{GD}$, $C_{SD}$, $C_{DG}$ are the intrinsic capacitances. Parasitic capacitances $C_{GSP}$, $C_{GDP}$, $C_{SDP}$, gate access resistance $R_G$, source and drain contact resistances $R_S$, $R_D$, respectively, are also depicted. Intrinsic model parameters are within the dashed box.

TABLE II

IV EXTRACTED PARAMETERS

| Parameter | Units | $L_g$=300 nm |
|---|---|---|
| $\mu$ | cm$^2$/(V·s) | 400 |
| $C_{back}$ | μF/cm$^2$ | 1.87 |
| $V_{BS0}$ | V | 1 |
| $R_c$ | Ω | 4 |
| $R_G$ | Ω | 37 |
| $\Delta$ | meV | 110 |
| $u_{sat}$ | m/s | 5.10$^6$ |

capacitance, $V_{BS0}$ the flat-band voltage, $R_c$-$R_G$ the contact and access gate resistances, $\Delta$ the inhomogeneity of the electrostatic potential, which is related to the residual charge density, and $u_{sat}$ the saturation velocity. Note that analogous $\mu$ values have been reported for similar GFETs in [20, 27]. A detailed methodology for the transport model parameters' extraction has been proposed in [32] while the very low $R_c$ value can be confirmed in [22, 30]. Notice that the rest of the parameters have been appropriately tuned to better fit the $\Re(Y_{21DEV})$ and $\Re(Y_{22DEV})$ experiments. While in [20] a precise physics-based transport model is presented which accounts separately for hole and electron contributions, this is not essential in the present study since experiments only from p-type region below Dirac voltage $V_{Dirac}$ are under discussion. Figure 2 depicts the real (left plots) and imaginary (right plots) parts of all the $Y$-parameters for four frequencies ($f$=2, 5, 10, 18 GHz) for $V_{DS}$=0.5 V and the model captures decently the de-embedded measured data for all the bias and frequency conditions, including $g_m$ and $g_{ds}$. An $R_G$=37 Ω alike [22] (for a similar GFET technology) is used in the Verilog-A simulations (cf. Table II) as it provides the best fitting for the measured $\Re(Y_{DEV})$. An overestimation of the experiments by the $\Im(Y_{22DEV})$ model is observed at $f$=10, 18 GHz at lower $V_{GS}$, probably caused by substrate coupling [33] (not considered in the model) as the thickness of the substrate $SiO_2/Si$ is around 300 nm/300 μm [30]. Dynamic substrate coupling effects in GFETs are out of the scope of this work. A |$g_m$| value of ~10 mS is recorded at the maximum measured $V_{GS}$=0.7 V which agrees with [22] for a similar GFET. The $g_m$ model is extended up to $V_{Dirac}$=1.2 V in the inset of figure 2a where |$g_m$| gets maximum at $V_{GS}$=0.9 V before starting to decrease steeply until it reaches 0 at the Dirac point [21].

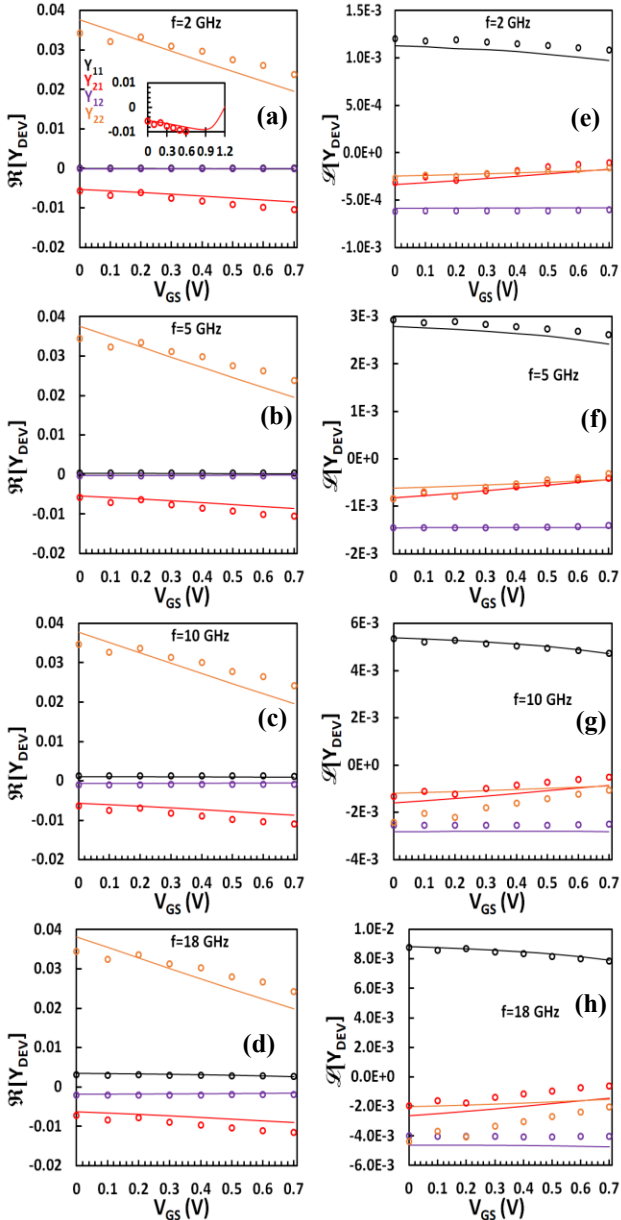

**Figure 2.** Real (left plots) and imaginary (right plots) part of de-embedded $Y$-parameters ($\Re[Y_{DEV}]$, $\mathscr{I}[Y_{DEV}]$) vs. gate voltage $V_{GS}$ with markers representing the measurements and lines the model for a GFET with gate width $W$=24 μm and length $L$=300 nm at different operation frequencies $f$=2 GHz (a, e), $f$=5 GHz (b, f), $f$=10 GHz (c, g) and $f$=18 GHz (d, h) at a drain voltage $V_{DS}$=0.5 V. $\Re[Y_{21DEV}]$ (=$g_m$) is shown in inset of (a) vs. $V_{GS}$ where the model is extended up to Dirac voltage $V_{Dirac}$=1.2 V.

To calculate the intrinsic QS small-signal parameters, $R_C$, $R_G$ must be removed from $Y_{DEV}$. For $R_C$, an advanced recently proposed method [23, 26-27], is applied to both the model and the experimental data and thus, $Y_{INT(RG)}$ yield [5, 28]:

$$
\begin{aligned}
&Y_{INT(RG)} \\
&= \begin{bmatrix} \omega^2 R_G C_{GG}^2 & \omega^2 R_G C_{GG} C_{GD} \\ g_{mi} - \omega^2 R_G C_{GG} C_{DG} & g_{dsi} + \omega^2 R_G C_{GG} C_{GD} \end{bmatrix} \\
&+ j \begin{bmatrix} \omega C_{GG} & -\omega C_{GD} \\ -\omega(C_{GD} + C_m + g_{mi} R_G C_{GG}) & \omega(C_{GD} + C_{SD} - g_{dsi} R_G C_{GG}) \end{bmatrix}
\end{aligned}
$$
$$(1a)$$

where $C_m$=$C_{DG}$-$C_{GD}$ is the gate transcapacitance accounting for the non-reciprocity of capacitances [5, 21], as mentioned earlier; $R_G$ contribution is still in the $Y_{INT(RG)}$. Intrinsic $Y$-parameters $Y_{INT}$ without $R_G$ effect are given by [21, 31]:

$$Y_{INT} = \begin{bmatrix} j\omega C_{GG} & -j\omega C_{GD} \\ g_{mi} - j\omega C_{DG} & g_{dsi} + j\omega(C_{GD} + C_{SD}) \end{bmatrix}$$
$$(1b)$$

From equations (1a), (1b), $C_{GG}$, $C_{GD}$, $C_{GS}$ can be extracted as:

$$C_{GG} = \frac{\mathscr{I}(Y_{11INT(RG)})}{\omega} = \frac{\mathscr{I}(Y_{11INT})}{\omega}, \qquad C_{GD} = -\frac{\mathscr{I}(Y_{12INT(RG)})}{\omega} = \frac{\mathscr{I}(Y_{12INT})}{\omega}, C_{GS} = C_{GG} - C_{GD}$$
$$(2)$$

where $\omega$ is the angular frequency. Intrinsic transconductance $g_{mi}$= $\Re(Y_{21INT})$ and $C_{DG}$ can be derived from:

$$\Re(Y_{21INT(RG)}) = g_{mi} - \omega^2 R_G C_{GG} C_{DG} \qquad (3a)$$
$$\mathscr{I}(Y_{21INT(RG)}) = -\omega(C_{DG} + g_{mi} R_G C_{GG}) \qquad (3b)$$

as all the other terms in equations (3a) and (3b) are known. Similarly, intrinsic output conductance $g_{dsi}$=$\Re(Y_{22INT})$ and $C_{SD}$ are estimated from:

$$\Re(Y_{22INT(RG)}) = g_{dsi} + \omega^2 R_G C_{GG} C_{GD} \qquad (4a)$$
$$\mathscr{I}(Y_{22INT(RG)}) = \omega(C_{GD} + C_{SD} - g_{dsi} R_G C_{GG}) \qquad (4b)$$

Thus, |$h_{21}$| and $U$ can be easily calculated as [21]:

$$|h_{21}(\omega)| = \left| -\frac{Y_{21}}{Y_{11}} \right| \rightarrow |h_{21}(2\pi f_t)| = 1 \qquad (5)$$
$$U(\omega) = -\frac{|Y_{12} - Y_{21}|^2}{4(\Re(Y_{11})\Re(Y_{22}) - \Re(Y_{12})\Re(Y_{21}))} \rightarrow U(2\pi f_{max}) = 1 \qquad (6)$$

and $f_t$, $f_{max}$ can also be derived as the frequencies where |$h_{21}$| and $U$ equal to unity (0 dB), respectively [4-5, 21], (cf. equations (5)-(6)). The explicit parameter extraction procedure described in the present Section is illustrated in the diagram of figure 3. The frequency-dependence of |$\Re(Y_{DEV(INT)})$| and |$\mathscr{I}(Y_{DEV(INT)})$|| is depicted in the main panels (insets) of figures 4a and 4b, respectively, at $V_{GS}$=0 V; $\Re(Y_{INT})$ and $\mathscr{I}(Y_{INT})$ are extracted from equations (1)-(4). $g_m$, $g_{ds}$ measurements are practically constant vs. $f$ with the models following this trend while an alike frequency-dependence is recorded for $g_{mi}$, $g_{dsi}$; $\Re(Y_{21INT})$=$\Re(Y_{11INT})$=0 according to equation (1b) thus, they are not included in the inset of figure 4a. $\mathscr{I}(Y_{DEV})$ experiments demonstrate an almost proportional-to-frequency behavior and they are successfully validated by the models, apart from $\mathscr{I}(Y_{22DEV})$ above 10 GHz, as mentioned earlier; an identical behavior is observed for $\mathscr{I}(Y_{INT})$.

Notice that $Y_{DEV}$ experiments directly reflect the frequency response of the device, as they are extracted after the de-embedding procedure before applying QS equations (1)-(4). Thus, the nearly proportional-to-frequency $\mathscr{I}(Y_{DEV})$ (cf. figure 4b) as well as the approximately constant-to-frequency $g_m$, $g_{ds}$ (cf. figure 4a) strongly indicate a QS regime of operation up to $f$=18 GHz [31]. Hence, charging resistances $R_{GS}$, $R_{GD}$ connected in series with $C_{GS}$, $C_{GD}$ to model NQS effects in [24, 31] (figure 8.20)], are ignored in the QS equivalent circuit in figure 1b because of the aforementioned $Y_{DEV}$ frequency-dependent relations. The latter simplifies the first-order NQS model proposed in [31] (equations (8.4.62), (8.4.69), (8.4.70), (8.4.72))] to the QS approach of the present study (cf. equations (1)-(4)). Measured $\Re(Y_{22DEV})$ fluctuates from the proportional to frequency dependence above 10 GHz (cf. figure 4b) and such behavior is associated with substrate coupling, as it has been already mentioned. There could be arguments that this trend is due to NQS effects at frequencies near $f_{tEXT}$ but in such case the rest of the experimental $\Re(Y_{DEV})$ parameters would have also been affected and since the latter is not the case as detailed before, NQS effects are discarded.

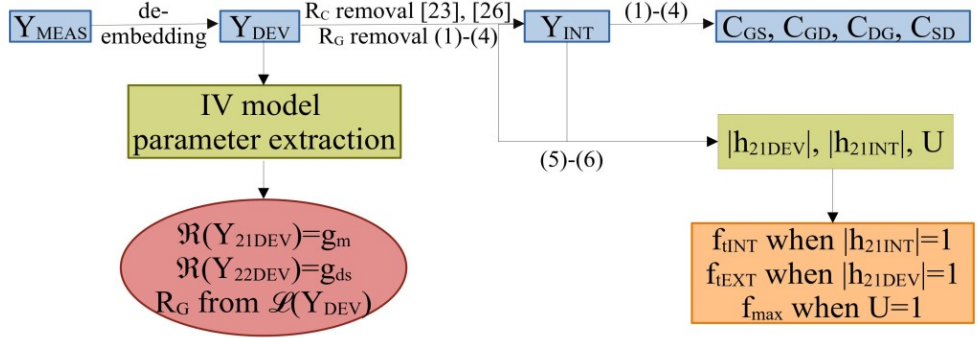

**Figure 3.** Small-signal parameter extraction flow chart

## 4. Results and Discussion

All the intrinsic capacitances in the equivalent circuit of figure 1 are fully characterized in the present analysis. Both modeled and measured capacitances are presented in figures 4c-4f, vs. $V_{GS}$ at four frequencies ($f$=2, 5, 10, 18 GHz) (apart from $C_{GS}$ which can be easily extracted from the last term of equation (2)). They are also shown vs. frequency at $V_{GS}$=0, 0.6 V in the two insets of figures 4c and 4d, respectively, to describe their frequency-dependence as well. A slight $V_{GS}$-dependence is recorded for measured $C_{GG}$, $C_{GD}$, $C_{DG}$ for a $V_{GS}$ below 0.7 V towards p-type region in agreement with findings in bibliography [20-22]. The intrinsic capacitance models extracted by the straightforward procedure described in equations (1)-(4), qualitatively capture this dependence. Regarding $C_{DG}$, which differs from $C_{GD}$ in contrast to a Meyer-like model as non-reciprocities are considered, such a $V_{GS}$-dependent model experimental validation is presented for the first time in GFETs. Notice in the insets of figures 4c and 4d, the weak frequency-dependence of measured $C_{GG}$, $C_{GD}$, $C_{DG}$ derived directly from the almost proportional-to-frequency $\Re(Y_{11INT(DEV)})$, $\Re(Y_{12INT(DEV)})$, $\Re(Y_{21INT(DEV)})$, respectively, (cf. figure 4b) through equations (2)-(3); the accuracy of the models for both $V_{GS}$ points is also remarkable. Recorded $C_{GG}$, $C_{GD}$ values around ~85-45 fF, respectively, are consistent with those referred in [22] for a similar GFET. These values in [22] are without $R_G$, $R_C$ removal but $R_C$ is very low for the specific GFET while $C_{GG}$, $C_{GD}$, are $R_G$-independent as is apparent from equation (2).

Similarly to $C_{DG}$, the $V_{GS}$- and frequency-dependence of a GFET $C_{SD}$ capacitance model is for the first time validated with experimental data and presented here. Measured $C_{SD}$ increases towards strong p-type regime at lower frequencies up to 5 GHz and the model follows this trend as depicted in figures 4c and 4d. For $f \gtrsim 10$ GHz in figures 4e and 4f, the model overestimates the experiments due to substrate coupling, following $\Re(Y_{22INT(DEV)})$ behavior (cf. figures 2g, 2h, 4b) as $C_{SD}$ and $\Re(Y_{22INT(DEV)})$ are strongly related through equations (1a)-(1b), (4a)-(4b). This can also be observed in the insets of figures 4c and 4d where $C_{SD}$ is depicted vs. frequency, where measured values start to decrease abruptly after $f \gtrsim 10$ GHz in contrast to the model which remains approximately fixed with frequency. Experimental $C_{SD}$ fluctuates from a maximum of ~55 fF at $V_{GS}$=0 V, $f$=2 GHz to a minimum of ~10 fF at $V_{GS}$=0.7 V, $f$=18 GHz and such values are much higher than those in [21] which is the only prior work where $C_{SD}$ experiments are reported. This is associated with an increased $R_G$=37 Ω (cf. Table II). The effect of $R_G$ on $C_{DG}$, $C_{SD}$ ($C_{GG}$, $C_{GD}$, $C_{GS}$ are $R_G$- independent as mentioned before), is illustrated in figure 5 where simulated intrinsic capacitances are shown vs. frequency at $V_{GS}$=0 V for the extracted value of $R_G$ as well as for a decreased one (~half). The application of a smaller $R_G$ value decreases the estimated $C_{SD}$ (~-5 fF); such negative experimental values are recorded in [21] also. Notice that $C_{DG}$ values for the low $R_G$ are also comparable with the measured values in [21] which, as in $C_{SD}$ case, is the only previous study that presents $C_{DG}$ measurements.

Critical RF figures of merit (FoMs) such as $f_{tINT}$, $f_{tEXT}$, $f_{max}$, $|h_{21}|$ and $U$ are also examined thoroughly. They are extracted from RF measurements and modeled in terms of bias and frequency. $|h_{21}|$, $U$ are inversely proportional to frequency [4, 20-21, 23] while the frequencies where they become equal to unity are defined as $f_t$ and $f_{max}$, respectively [4, 21] (cf. equations (5)-(6)). Notice that if $|h_{21}|$ in equation (5) is extracted from $Y_{DEV}$ ($|h_{21DEV}|$) where $R_C$, $R_G$ are still considered, $f_{tEXT}$ can be derived while if $|h_{21}|$ is calculated from $Y_{INT}$ ($|h_{21INT}|$) after $R_C$, $R_G$ elimination then $f_{tINT}$ is estimated. That said, $|h_{21DEV}|$, $f_{tINT}$ depend directly on $g_{mi}$ while $|h_{21DEV}|$, $f_{EXT}$ on $g_m$, respectively. Extrinsic $U$ and consequently, extrinsic $f_{max}$ can be extracted by equation (6) if $Y_{DEV}$ are considered. Simulated $f_{tINT}$, $f_{tEXT}$ in figure 6a and $f_{max}$ in figure 6b are precisely validated with measurements vs. $V_{GS}$ at $V_{DS}=0.5$ V; the models are extended up to $V_{Dirac}=1.2$ V and also presented for $V_{DS}=0.1$ V (red lines) and for $L=100$ nm at $V_{DS}=0.5$ V (green lines). The small $R_C$ value for the GFET under test accounts for the trivial degradation of $f_{tEXT}$ in comparison with $f_{tINT}$. Both measured and simulated $f_t$, $f_{max}$ increase with $V_{GS}$ similarly to $|g_m|$ (cf. figure 2a), whereas, models continue to increase up to $V_{GS}=0.9$ V and then fall abruptly towards Dirac point [20-21]. $f_{tEXT}$, $f_{max}$ values around 19 and 12 GHz, respectively, are extracted at the maximum measured $V_{GS}=0.7$ V which agree with a similar GFET in [22]; operation frequencies above $f_{tEXT}$ would induce NQS effects which are beyond the scope of the present study. Simulated $f_{tEXT}$, $f_{tINT}$ and $f_{max}$ considerably increase for the smaller $L$ and significantly decrease for the lower $V_{DS}$ case, respectively, confirming previous experimental findings [3-5]. The insets in figures 6a and 6b depict experimental $|h_{21DEV}|$, $U$, respectively, vs. frequency at $V_{GS}=0$ V, $V_{DS}=0.5$ V where their inversely proportional relation with frequency is well described by the models.

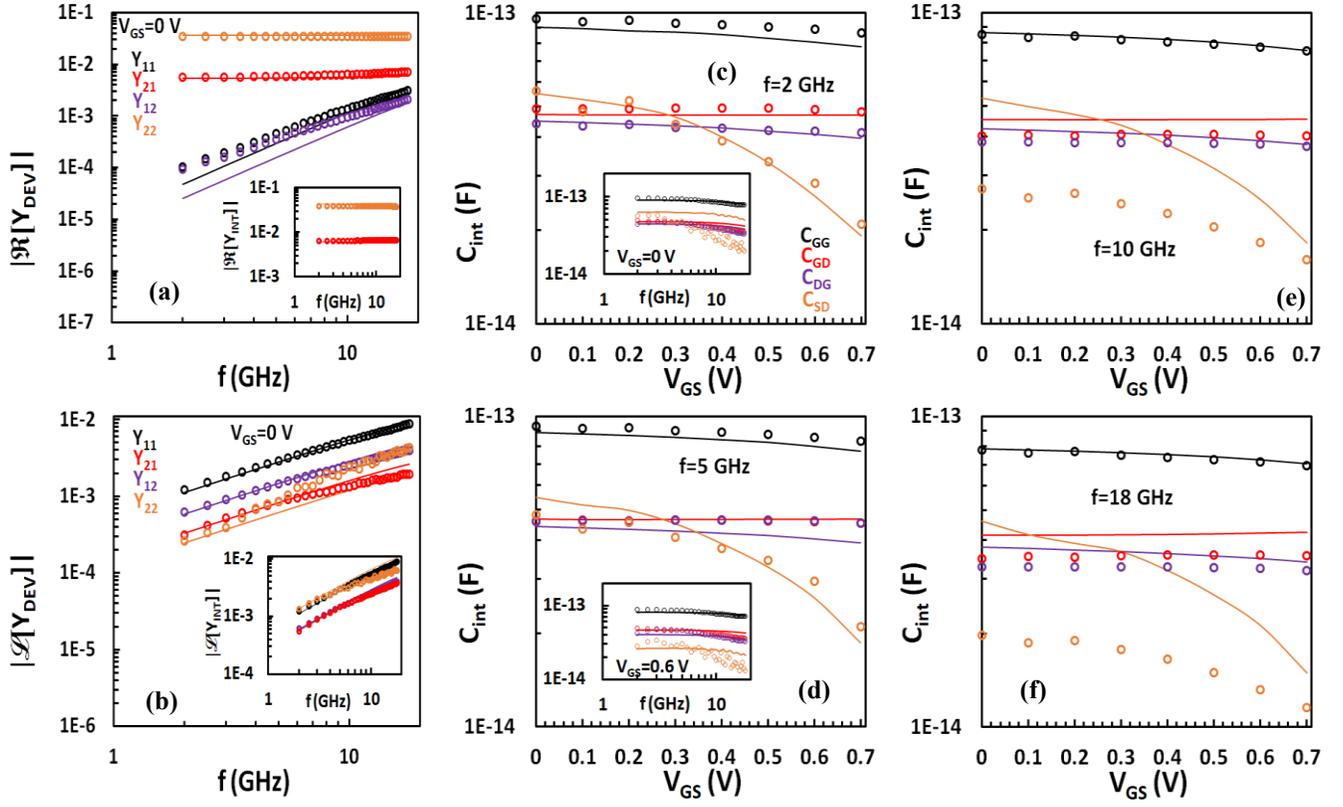

**Figure 4.** (a) $|\Re[Y_{DEV}]|$ ($|\Re[Y_{INT}]|$ in inset), (b) $|\mathcal{I}[Y_{DEV}]|$ ($|\mathcal{I}[Y_{INT}]|$ in inset) vs. $f$ at $V_{GS}=0$ V and intrinsic capacitances $C_{GG}$, $C_{GD}$, $C_{DG}$, $C_{SD}$, respectively vs. $V_{GS}$ at different operation frequencies $f=2$ GHz (c), $f=5$ GHz (d), $f=10$ GHz (e) and $f=18$ GHz (f) and vs. $f$ in insets for $V_{GS}=0$ (c), 0.6 V (d) for a GFET with $W=24$ μm and $L=300$ nm at $V_{DS}=0.5$ V. Markers represent the measurements and lines the models.

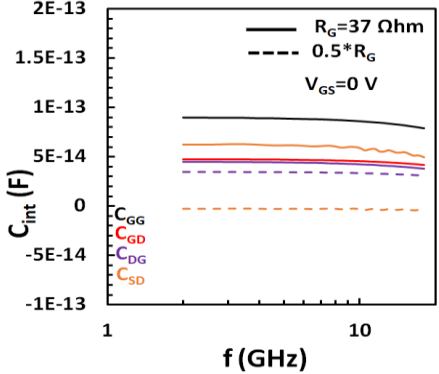

**Figure 5.** Simulated $C_{GG}$, $C_{GD}$, $C_{DG}$, $C_{SD}$, respectively, vs. $f$ for $V_{GS}=0$ V for a GFET with $W=24$ μm and $L=300$ nm at $V_{DS}=0.5$ V. Solid lines: Extracted $R_G=37$ Ω, dashed lines: $R_G$ value decreased by a factor of 2.

There are no studies presenting the bias-dependence of $|h_{21}|$, $U$ FoMs in GFETs and this is accomplished in the present work. Both experimental and simulated $|h_{21DEV}|$ and $|h_{21INT}|$ in figure 7a as well as $U$ in figure 7b are shown vs. $V_{GS}$ at $V_{DS}=0.5$ V for four operating frequencies ($f=2$, 5, 10, 18 GHz); the models are extended up to $V_{Dirac}=1.2$ V for the 2 GHz case and also depicted for $V_{DS}=0.1$ V (blue lines) and for $L=100$ nm at $V_{DS}=0.5$ V (green lines) for the aforementioned frequency. The models account well for the measured data at any bias and frequency point while both $|h_{21}|$, $U$ present a similar trend vs. $V_{GS}$ as $f_t$, $f_{max}$ and $|g_m|$. Maximum measured $|h_{21DEV}|$, $|h_{21INT}|$, $U$ are placed at $V_{GS}=0.7$ V, $f=2$ GHz while simulations rise up to $V_{GS}=0.9$ V before starting to decrease steeply, similarly with $f_t$, $f_{max}$ and $|g_m|$. Finally, modeled $|h_{21DEV}|$, $|h_{21INT}|$ and $U$ are quite reduced at the lower $V_{DS}$ agreeing with experiments in [20], while they are heightened at the smaller $L$.

In general, a decent percentage error is recorded between experiments and models from 0.3% in the best case to 10%

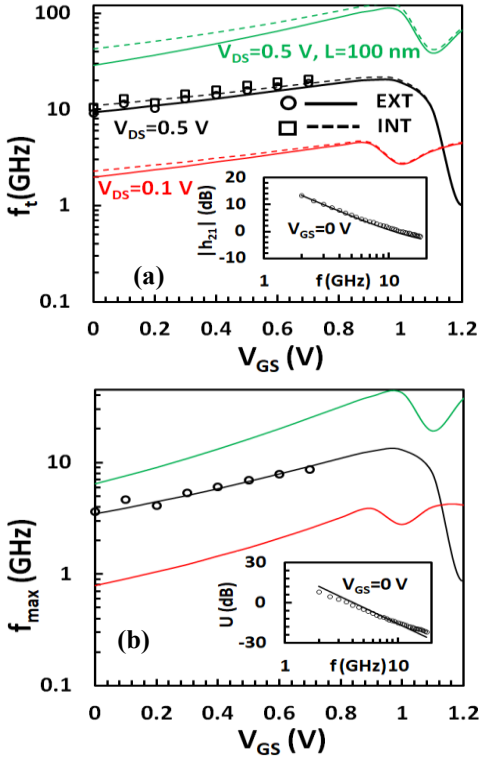

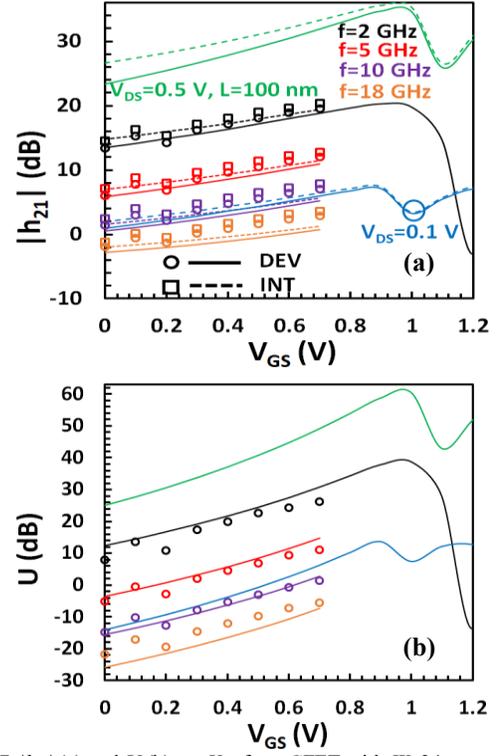

**Figure. 6.** Intrinsic and extrinsic cut-off frequencies $f_{tINT}$, $f_{tEXT}$, respectively, (a) and extrinsic maximum oscillation frequency $f_{max}$ (b) vs. $V_{GS}$, small-signal current gain $|h_{21}|$ in (a) inset and unilateral power gain $U$ in (b) inset vs. $f$ for $V_{GS}$=0 V for a GFET with $W$=24 µm and $L$=300 nm (green lines: $L$=100 nm) at $V_{DS}$=0.5 V (black lines) and $V_{DS}$=0.1 V (red lines). Markers represent the measurements and lines the model; dashed lines: $f_{tINT}$ model.

**Figure 7.** $|h_{21}|$ (a) and $U$ (b) vs. $V_{GS}$ for a GFET with $W$=24 µm and $L$=300 nm at different operation frequencies $f$= 2, 5, 10, 18 GHz (green lines: $L$=100 nm, $f$=2 GHz) at $V_{DS}$=0.5 V. Markers represent the measurements and lines the model; dashed lines: $|h_{21INT}|$ model. Models are also shown for $V_{DS}$=0.1 V at $f$= 2 GHz with blue lines.

in the worst case for most of the parameters investigated. The latter strengthen the validity of the present study, especially after considering that more than 10 small-signal parameters are extracted simultaneously in terms of both their bias- and frequency-dependence. Regarding $C_{SD}$, the error remains low around 5% below $f$=5 GHz but it considerably increases at 10 and 18 GHz, respectively, as it is expected due to substrate coupling contribution which is not yet included in the model. Experimental discrepancies are also recorded for $U$ at frequencies near $f_{tEXT}$ in agreement with bibliography for both MOSFETs [28] and GFETs [20], resulting in higher errors and less accuracy of the model in this regime.

## 5. Conclusion

A charge-based small-signal GFET model and a parameter extraction methodology have been presented and validated with measurements from a 300 nm RF GFET for several $V_{GS}$ values up to $f$=18 GHz. Explicit efficient methods are applied for the removal of $R_c$ and $R_G$. As a result, straightforward expressions for small-signal parameters and most significant RF FoMs have been obtained. Despite the high computational power available almost everywhere nowadays, such direct and explicit derivations are critical for an initial acceptable estimation of model parameters, which is crucial from circuit design aspect. On the contrary, complex procedures for $R_c$ and $R_G$ elimination lead to complicated mathematics [21] which are hard to be handled for fast circuit-analysis. Moreover, optimization routines deployed for the extraction of a plethora of parameters for compact modeling purposes of complete transistor technologies (tens or hundreds on-wafer devices with different footprints and a broad range of bias-points) result in more accurate results if a sufficient first estimation of the parameters is provided to the algorithm. Such calculations can be accomplished with methodologies similar to the one proposed in the present study.

The successful experimental validation of the frequency-dependence of all GFET intrinsic capacitance models has not been shown elsewhere. Besides, both bias- and frequency-dependent $C_{SD}$, $C_{DG}$ capacitance models are for the first time demonstrated and validated accurately with experiments. Note that $C_{DG}$ is considered equal to $C_{GD}$ in an inaccurate Meyer-like approach which is not the case in this work, while $C_{SD}$ had almost always been neglected in relevant bibliography, so far. The effect of $R_G$ on $C_{SD}$, $C_{DG}$ is also highlighted. Finally, the accurate $V_{GS}$ behavior of $|h_{21}|$, $U$ models in comparison with experiments is also presented for the first time and reveals a strong relation with $f_t$, $f_{max}$ trend vs. $V_{GS}$.

## Acknowledgements

This work has received funding from the European Union's Horizon 2020 research and innovation programme under grant agreements No GrapheneCore3 881603, from Ministerio de Ciencia, Innovación y Universidades under grant agreements RTI2018-097876-B-C21(MCIU/AEI/FEDER, UE), FJC2020-046213-I and PID2021-127840NB-I00 (MCIN/AEI/FEDER, UE) and by the European Union Regional Development Fund within the framework of the ERDF Operational Program of Catalonia 2014-2020 with the support of the Department de Recerca i Universitat, with a grant of 50% of total cost eligible. GraphCAT project reference: 001-P-001702.

## References

[1] Saeed M, Palacios P, Wei M-D, Baskent E, Fan C-Y, Uzlu B, Wang K-T, Hemmetter A, Wang Z, Neumaier D, Lemme M C, Negra R. Graphene-Based Microwave Circuits: A Review. Advanced Materials 2021. https://doi.org/10.1002/adma.202108473.

[2] Wu Y, Zou X, Sun M, Cao Z, Wang X, Huo S, Zhou J, Yang Y, Yu X, Kong Y, Yu G, Liao L, Chen T. 200 GHz Maximum Oscillation Frequency in CVD Graphene Radio Frequency Transistors. ACS Applied Materials and Interfaces 2016; 8(39):25645. https://pubs.acs.org/doi/abs/10.1021/acsami.6b05791.

[3] Bonmann M, Asad M, Yang X, Generalov A, Vorobiev A, Banszerus L, Stampfer C, Otto M, Neumaier D, Stake J. Graphene field-effect transistors with high extrinsic fT and fmax. IEEE Electron Device Letters 2019; 40(1):131-134. https://doi.org/10.1109/LED.2018.2884054.

[4] Asad M, Jeppson K O, Vorobiev A, Bonmann M, Stake J. Enhanced High-Frequency Performance of Top-Gated Graphene FETs Due to Substrate-Induced Improvements in Charge Carrier Saturation Velocity. IEEE Transactions on Electron Devices 2021; 68(2):899-902. https://doi.org/10.1109/TED.2020.3046172.

[5] Enz C. An MOS transistor model for RF IC design valid in all regions of operation. IEEE Transactions on Microwave Theory and Techniques 2002; 50(1):342-359. https://doi.org/10.1109/22.981286.

[6] Antonopoulos A, Bucher M, Papathanasiou K, Mavredakis N, Makris N, Sharma R K, Sakalas P, Schroter M. CMOS small-signal and thermal noise modeling at high frequencies. IEEE Transactions on Electron Devices 2013; 60(11):3726-3733. https://doi.org/10.1109/TED.2013.2283511.

[7] Hadarig A I, Hoeye S ver, Fernández M, Vázquez C, Alonso L, Las-Heras F. 330-500 THz graphene-based single-stage high-order subharmonic mixer. IEEE Access 2019; 7:113151-113160. https://doi.org/10.1109/ACCESS.2019.2935310.

[8] Hamed A, Saeed M, Negra R. Graphene-Based Frequency-Conversion Mixers for High-Frequency Applications. IEEE Transactions on Microwave Theory and Techniques 2020; 68(6):2090-2096. https://doi.org/10.1109/TMTT.2020.2978821.

[9] Yu C, He Z, Song X, Gao X, Liu Q, Zhang Y, Yu G, Han T, Liu C, Feng Z, Cai S. Field Effect Transistors and Low Noise Amplifier MMICs of Monolayer Graphene. IEEE Electron Device Letters 2021; 42(2):268-271. https://doi.org/10.1109/LED.2020.3045710.

[10] Hamed A, Asad M, Wei M-D, Vorobiev A, Stake J, Negra R. Integrated 10-GHz Graphene FET Amplifier. IEEE Journal of Microwaves 2021; 1(3):821-826. https://doi.org/10.1109/JMW.2021.3089356.

[11] Kabir H M D, Salahuddin S M. A frequency multiplier using three ambipolar graphene transistors. Microelectronics Journal 2017; 70:12-15. https://doi.org/10.1016/j.mejo.2017.10.002.

[12] Hamed A, Habibpour O, Saeed M, Zirath H, Negra R. W-Band Graphene-Based Six-Port Receiver. IEEE Microwave and Wireless Components Letters 2018; 28(4):347-349. https://doi.org/10.1109/LMWC.2018.2808416.

[13] Fadil D, Passi V, Wei W, Salk S ben, Zhou D, Strupinski W, Lemme M C, Zimmer T, Pallecchi E, Happy H, Fregonese S. A broadband active microwave monolithically integrated circuit balun in graphene technology. Applied Sciences (Switzerland) 2020; 10(6):2183. https://doi.org/10.3390/app10062183.

[14] Thiele S A, Schaefer J A, Schwierz F. Modeling of graphene metal-oxide-semiconductor field-effect transistors with gapless large-area graphene channels. Journal of Applied Physics 2010; 107:094505. https://doi.org/10.1063/1.3357398.

[15] Champlain J G. A physics-based, small-signal model for graphene field effect transistors. Solid-State Electronics 2012; 67(1):53-62. https://doi.org/10.1016/j.sse.2011.07.015.

[16] Habibpour O, Vukusic J, Stake J. A large-signal graphene FET model. IEEE Transactions on Electron Devices 2012; 59(4):968-975. https://doi.org/10.1109/TED.2012.2182675.

[17] Fregonese S, Magallo M, Maneux C, Happy H, Zimmer T. Scalable electrical compact modeling for graphene FET transistors. IEEE Transactions on Nanotechnology 2013; 12(4):539-546. https://doi.org/10.1109/TNANO.2013.2257832.

[18] Rodriguez S, Vaziri S, Smith A, Fregonese S, Ostling M, Lemme M C, Rusu A. A comprehensive graphene FET model for circuit design. IEEE Transactions on Electron Devices 2014; 61(4):1199-1206. https://doi.org/10.1109/TED.2014.2302372.

[19] Pasadas F, Jiménez D. Large-Signal Model of Graphene Field-Effect Transistors - Part I: Compact Modeling of GFET Intrinsic Capacitances. IEEE Transactions on Electron Devices 2016; 63(7):2936-2941. https://doi.org/10.1109/TED.2016.2570426.

[20] Aguirre-Morales J D, Fregonese S, Mukherjee C, Wei W, Happy H, Maneux C, Zimmer T. A Large-Signal Monolayer Graphene Field-Effect Transistor Compact Model for RF-Circuit Applications. IEEE Transactions on Electron Devices 2017; 64(10):4302-4309. https://doi.org/10.1109/TED.2017.2736444.

[21] Pasadas F, Wei W, Pallecchi E, Happy H, Jiménez D. Small-Signal Model for 2D-Material Based FETs Targeting Radio-Frequency Applications: The Importance of Considering Nonreciprocal Capacitances. IEEE Transactions on Electron Devices 2017; 64(11):4715-4723. https://doi-org.are.uab.cat/10.1109/TED.2017.2749503.

[22] Deng M, Fadil D, Wei W, Pallecchi E, Happy H, Dambrine G, de Matos M, Zimmer T, Fregonese S. High-Frequency Noise Characterization and Modeling of Graphene Field-Effect Transistors. IEEE Transactions on Microwave Theory and Techniques 2020; 68(6):2116-2123. https://doi-org.are.uab.cat/10.1109/TMTT.2020.2982396.

[23] Pacheco-Sanchez A, Ramos-Silva J N, Ramirez-Garcia E, Jimenez D. A Small-Signal GFET Equivalent Circuit Considering an Explicit Contribution of Contact Resistances. IEEE Microwave and Wireless Components Letters 2021; 31(1):29-32. https://doi.org/10.1109/LMWC.2020.3036845.

[24] Pasadas F, Jiménez D. Non-Quasi-Static Effects in Graphene Field-Effect Transistors Under High-Frequency Operation. IEEE Transactions on Electron Devices 2017; 67(5):2188-2196. https://doi-org.are.uab.cat/10.1109/TED.2020.2982840.

[25] MEYER J E 1971 MOS MODELS AND CIRCUIT SIMULATION R.C.A. Review 32.

[26] Ramos-Silva J N, Pacheco-Sánchez A, Enciso-Aguilar M A, Jiménez D, Ramírez-García E. Small-signal parameters extraction and noise analysis of CNTFETs. Semiconductor Science and Technology 2020; 35(4):045024. https://doi.org/10.1088/1361-6641/ab760b.

[27] Mavredakis N, Pacheco-Sanchez A, Sakalas P, Wei W, Pallecchi E, Happy H, Jimenez D. Bias-dependent intrinsic RF thermal noise modeling and characterization of single-layer graphene FETs. IEEE Transactions on Microwave Theory and Techniques 2021; 69(11):4639-4646. https://doi.org/10.1109/TMTT.2021.3105672.

[28] Chalkiadaki M A, Enz C C. RF Small-Signal and Noise Modeling Including Parameter Extraction of Nanoscale MOSFET From Weak to Strong Inversion. IEEE Transactions on Microwave Theory and Techniques 2015; 63(7):2173-2184. https://doi.org/10.1109/TMTT.2015.2429636.

[29] Wei W, Zhou X, Deokar G, Kim H, Belhaj M M, Galopin E, Pallecchi E, Vignaud D, Happy H. Graphene FETs with aluminum bottom-gate electrodes and its natural oxide as dielectrics. IEEE Transactions on Electron Devices 2015; 62(9):2769-2773. https://doi.org/10.1109/TED.2015.2459657.

[30] Wei W, Fadil D, Pallecchi E, Dambrine G, Happy H, Deng M, Fregonese S, Zimmer T. High frequency and noise performance of GFETs, in Proc IEEE International Conference on Noise and Fluctuations (ICNF). Vilnius, Lithuania; Jun. 2017. https://doi.org/10.1109/ICNF.2017.7985969.

[31] Tsividis Y, McAndrew C. 2011 Operation and Modeling of the MOS Transistor (3$^{rd}$ edition, New York: Oxford University Press).

[32] Mavredakis N, Wei W, Pallecchi E, Vignaud D, Happy H, Cortadella R G, Schaefer N, Calia A B, Garrido J A, Jimenez D. Low-frequency noise parameter extraction method for single-layer graphene FETs. IEEE Transactions on Electron Devices 2020; 67(5):2093-2099. https://doi.org/10.1109/TED.2020.2978215.

[33] Cao Y, Zhang W, Fu J, Wang Q, Liu L, Guo A. A Novel Parameter Extraction Technique of Microwave Small-Signal Model for Nanometer MOSFETS. IEEE Microwave and Wireless Components Letters 2019; 29(11):710-713. https://doi.org/10.1109/LMWC.2019.2942193.